\documentclass[sigconf]{acmart}
\AtBeginDocument{%
  \providecommand\BibTeX{{%
    \normalfont B\kern-0.5em{\scshape i\kern-0.25em b}\kern-0.8em\TeX}}}

\setcopyright{acmcopyright}
\copyrightyear{2021}
\acmYear{2021}
\acmDOI{10.1145/1122445.1122456}

\acmConference[WWW '22]{WWW '22: ACM Web Conference 2022}{April 25--29, 2022}{Lyon, France}
\acmBooktitle{WWW '22: ACM Web Conference}
\acmPrice{15.00}
\acmISBN{978-1-4503-XXXX-X/18/06}

\usepackage{booktabs}
\usepackage{threeparttable}
\usepackage{float}
\usepackage{multirow}
\usepackage{textcomp}
\usepackage{caption}
\usepackage{booktabs}
\usepackage{bm}
\usepackage{algorithm}
\usepackage{algorithmic}
\usepackage{makecell}
\usepackage{url}
\usepackage[labelformat=simple]{subcaption}
\usepackage{CJKutf8}

\usepackage{enumitem}

\newcommand{\bC}{\mathbf{C}}

\newcommand{\bW}{\mathbf{W}}

\newcommand{\bc}{\mathbf{c}}
\newcommand{\bd}{\mathbf{d}}

\newcommand{\bv}{\mathbf{v}}

\newcommand{\ie}{\textit{i.e.}}

\newcommand{\minitab}[2][l]{\begin{tabular}{#1}#2\end{tabular}}

\copyrightyear{2022}
\acmYear{2022}
\setcopyright{acmlicensed}\acmConference[WWW '22]{Proceedings of the ACM Web Conference 2022}{April 25--29, 2022}{Virtual Event, Lyon, France}
\acmBooktitle{Proceedings of the ACM Web Conference 2022 (WWW '22), April 25--29, 2022, Virtual Event, Lyon, France}
\acmPrice{15.00}
\acmDOI{10.1145/3485447.3511962}
\acmISBN{978-1-4503-9096-5/22/04}

\begin{document}
\fancyhead{}

\title{Socialformer: Social Network Inspired Long Document Modeling for Document Ranking}

\author{Yujia Zhou$^{2}$, Zhicheng Dou$^{1,4*}$, Huaying Yuan$^{3}$, and Zhengyi Ma$^{2}$}
\affiliation{%
  $^1$Gaoling School of Artificial Intelligence, Renmin University of China \\
  $^2$School of Information, Renmin University of China \\
  $^3$College of Computer Science, Nankai University \\
  $^4$Beijing Key Laboratory of Big Data Management and Analysis Methods \\
 \country{}
}
\email{{zhouyujia, *dou}@ruc.edu.cn}
\def\authors{Yujia Zhou, Zhicheng Dou, Huaying Yuan, and Zhengyi Ma}
\settopmatter{printacmref=true}
\begin{abstract}
Utilizing pre-trained language models has achieved great success for neural document ranking. Limited by the computational and memory requirements, long document modeling becomes a critical issue. Recent works propose to modify the full attention matrix in Transformer by designing sparse attention patterns. However, most of them only focus on local connections of terms within a fixed-size window. How to build suitable remote connections between terms to  better model document representation remains underexplored. In this paper, we propose the model Socialformer, which introduces the characteristics of social networks into designing sparse attention patterns for long document modeling in document ranking. Specifically, we consider several attention patterns to construct a graph like social networks. Endowed with the characteristic of social networks, most pairs of nodes in such a graph can reach with a short path while ensuring the sparsity. To facilitate efficient calculation, we segment the graph into multiple subgraphs to simulate friend circles in social scenarios. Experimental results confirm the effectiveness of our model on long document modeling.
\end{abstract}



\maketitle

\section{Introduction}\label{sec:intro}
Document ranking is a crucial task in information retrieval. It focuses on generating an ordered document list in response to the user's query. In recent years, pre-trained language models, such as BERT~\cite{bert}, have made impressive progress in natural language processing and information retrieval. Its powerful contextual representation learning ability is suitable for modeling documents in semantic space and has been widely applied in information retrieval \cite{bertmaxp, bertanalyze,DBLP:conf/cikm/MaDXZJCW21}. However, due to the memory constraints of quadratic attention matrix, the input sequence of the BERT-based model is limited to 512 tokens~\cite{DBLP:conf/emnlp/AinslieOACFPRSW20}. Therefore, how to apply BERT over long documents remains a challenge for document ranking.

To deal with this problem, some early studies~\cite{idcm, bertmaxp, bertanalyze} divide a document into multiple passages with fixed-size windows, match the query with each passage, and aggregate passage-level relevance signals for document ranking. These works concentrate on the semantic modeling in each passage, but ignore the word level interactions between passages. This prevents the model from learning a global document representation. Subsequently, another group of works proposes handling long document with sparse attention patterns in Transformer~\cite{DBLP:conf/nips/VaswaniSPUJGKP17}, such as sliding window attention~\cite{poolingformer, longformer, bigbird, tkl}, dilated window attention~\cite{sparsetransformer, longformer}, and global attention~\cite{longformer, bigbird}. These patterns enlarge the receptive field of each term to interact with more distant terms while ensuring the sparsity.

Previous methods have made great progress in reducing complexity of self-attention layer with sparse connections. Most of them mainly concentrate on building local connections of terms to model semantic dependencies inside a fixed-size sliding window. However, in these methods, remote connections between terms are ignored or captured by simple patterns~\cite{longformer, bigbird}. In fact, based on small-world theory~\cite{watts1998collective,kleinberg2000navigation}, suitable remote connections in a sparse social network can shorten the path between most pairs of nodes. Bigbird~\cite{bigbird} first introduces the small-world graph in building remote connections with random attention. However, in the real social network, the formation of remote connection between people is not random, but is related to the distance between them and their status in social networks. Inspired by it, \textbf{we attempt to leverage the characteristics of social networks to build well designed remote connections of terms within a long document}. With the social network inspired graph, we are able to achieve better information propagation ability and finally yield better document representations for Web document ranking.


Social networks have been thoroughly investigated by many researchers. There are three main characteristics to ensure the effective transmission of information~\cite{holtzman2004social, freeman2004development, scott1988social}. \textbf{(1) Randomness}. Any two people have a certain probability to establish a new contact. \textbf{(2) Distance-aware}. In a small-world network, the probability of constructing connection between two people should follow the inverse square law with distance~\cite{kleinberg2000navigation}. \textbf{(3) Centrality}. Some celebrities possess more influence over social networks and are usually connected with more people. These characteristics ensure that even if the graph is sparse, most pairs of nodes can reach with a short path. This will assure the efficient information exchange over the sparse social network. Inspired by these characteristics, \textbf{we propose a similar paradigm to form the sparse attention matrix in long document modeling}. Different from traditional fixed attention patterns, all connections between words are sampled according to the probability. This enables us to dynamically adjust the edge distribution based on the document length and content. The calculation of the probability follows the characteristics of social networks, which take the word distance and word centrality into account. Under such a strategy, the graph we construct imitating social networks can enhance the information transmission in the document while ensuring the sparsity.

Due to the randomness of our probability sampled graph, how to achieve fast calculation becomes a new challenge. The reason why previous sparse attention matrices can handle long documents is that they can be easily split into multiple small self-attention blocks. To facilitate calculation, \textbf{we propose to segment the graph into multiple subgraphs}, while retaining as much information as possible. In social networks, the relationships between people usually depend on friend circles~\cite{henning1996strong, easley2012networks}, and there is often a central person in each circle~\cite{DBLP:conf/sigir/ZhouDWXW21}. Based on this observation, we propose a graph partition algorithm which focuses on finding central nodes in the graph. According to these central nodes, we are able to form multiple subgraphs simulating the friend circles in social scenarios. 

In general, members within a friend circle are connected through strong ties with frequent interactions~\cite{hu2019strong, henning1996strong}. By contrast, the interactions between friend circles through weak ties are relatively infrequent~\cite{wilson1998weak}. To model such interactions, we propose a two-stage method for information transmission. At the first stage, the \textbf{intra-circle interaction} is applied to model semantic dependency between terms within each subgraph. Second, for information transmission between circles, we carry out the \textbf{inter-circle interaction} on central nodes of multiple subgraphs. Under such a strategy, most pairs of words in the document can transmit information via a direct connection or multiple iterative stacking blocks.

More specifically, we propose Socialformer, a social network inspired long document modeling method for document ranking. Socialformer is composed of four steps. \textbf{First}, based on the characteristics of social networks, we design four sparse attention patterns to construct a graph with probability sampling. \textbf{Second}, we present two friend circle based strategies of graph partition to reduce the memory and computational complexity. \textbf{Third}, we devise a two-stage information transmission model to capture the interactions between terms with the augmented transformer structure. \textbf{Finally}, by aggregating the representations of passages and subgraphs, a comprehensive document representation is formed for document ranking. We conduct experiments on the widely used TREC DL and MS MARCO dataset~\cite{DBLP:journals/corr/abs-2003-07820} for document ranking. Experimental results show that our proposed model Socialformer significantly outperforms existing document ranking models.

The contributions are summarized as follows. (1) We introduce the social network theories into long document modeling, which provides a theoretical basis for enhancing information transmission in long documents. (2) Inspired by the characteristics of social networks, we devise several social-aware sparse attention patterns to build the graph with probability sampling. (3) To reduce complexity, a graph partition algorithm is proposed referring to the concept of friend circles in social networks. (4) We apply a two-stage information transmission model to achieve intra-circle and inter-circle interactions with the augmented transformer.

\section{Related Work}\label{sec:related_work}
\textbf{Passage-level Document Ranking}. A major disadvantage of the Transformer~\cite{DBLP:conf/nips/VaswaniSPUJGKP17} based models is that they cannot handle long documents due to the quadratic memory complexity. Inspired by using passage-level evidence for document ranking~\cite{DBLP:conf/sigir/Callan94}, an intuitive idea is to segment the document text into multiple small chunks, compare the query to all passages~\cite{DBLP:conf/cikm/RudraA20, DBLP:conf/ecir/AiOC18, bertmaxp, DBLP:conf/sigir/MacAvaneyYCG19, DBLP:journals/corr/abs-1901-04085, idcm}, then aggregate the information of each passage~\cite{parade, DBLP:conf/emnlp/YilmazWYZL19}. Some early studies focus on combining passage ranking scores with different strategies. \citet{bertmaxp} devised three ways (MaxP, FirstP, SumP) to get the document ranking scores from all passage-level scores. \citet{idcm} proposed an intra-document cascade ranking model with knowledge distillation to speed up selecting passages. However, these methods ignore the information transfer between passages. This will prevent the model from learning a global document representation. To deal with this problem, several representation aggregation methods were proposed to learn the global document embeddings. \citet{DBLP:conf/www/WuMLZZZM20} used LSTM~\cite{DBLP:journals/neco/HochreiterS97} to model the sequential information hidden in passages. \citet{parade} tried a series of representation aggregation strategies with max pooling, attention pooling, transformer, etc. In order to further strengthen the information transfer between passages, some hierarchical transformer structures were proposed to model intra-passage and inter-passage interactions~\cite{hibert, hiTransformer, DBLP:conf/naacl/YangYDHSH16, DBLP:conf/cikm/Yang00BN20}. To further learn the document embedding with a global view, some studies use iterative attention blocks to enlarge the receptive field of each term layer by layer, such as Transformer-XL~\cite{transformerxl} and Transformer-XH~\cite{transformerxh}. Although these efforts have achieved certain success, how to guide information propagation in a reasonable manner still remains underexplored.

\textbf{Long-Document Transformers}. Another idea to solve the problem of long document representation is to design sparse attention patterns~\cite{routingtransformer, longformer, reformer, informer}, so as to avoid computing the full quadratic attention matrix multiplication. One of the most intuitive attention patterns is the sliding window attention~\cite{poolingformer, longformer, bigbird, tkl, blockbert}, which only keeps links to surrounding terms. Moreover, dilated sliding window~\cite{longformer, poolingformer, sparsetransformer} was devised to further increase the receptive field without additional computational costs. To fit the specific tasks, several works proposed to use the global attention~\cite{longformer, bigbird, DBLP:conf/emnlp/AinslieOACFPRSW20, DBLP:journals/corr/abs-2006-03274} to highlight the influence of certain tokens. In the field of information retrieval, query terms are usually set as global tokens to attend to all tokens~\cite{qdstransformer}. To model the document structure, some graph-based transformer methods~\cite{pgt, graphormer} were presented to lower computational costs. However, any two words in the document should have a probability to be connected~\cite{watts1998collective}. To implement this idea, \citet{bigbird} applied random attention to construct the sparse attention matrix, which brought significant improvement compared to structured patterns. In this paper, we integrate the social network theory to build remote edges for long document modeling. Our model can enhance the information transmission and learn comprehensive representations for document ranking.

\section{Methodology}\label{sec:model}
\begin{figure*}[!t]
	\centering
	\vspace{-0.2cm}
	\setlength{\abovecaptionskip}{0.1cm}
	\includegraphics[width=0.75\linewidth]{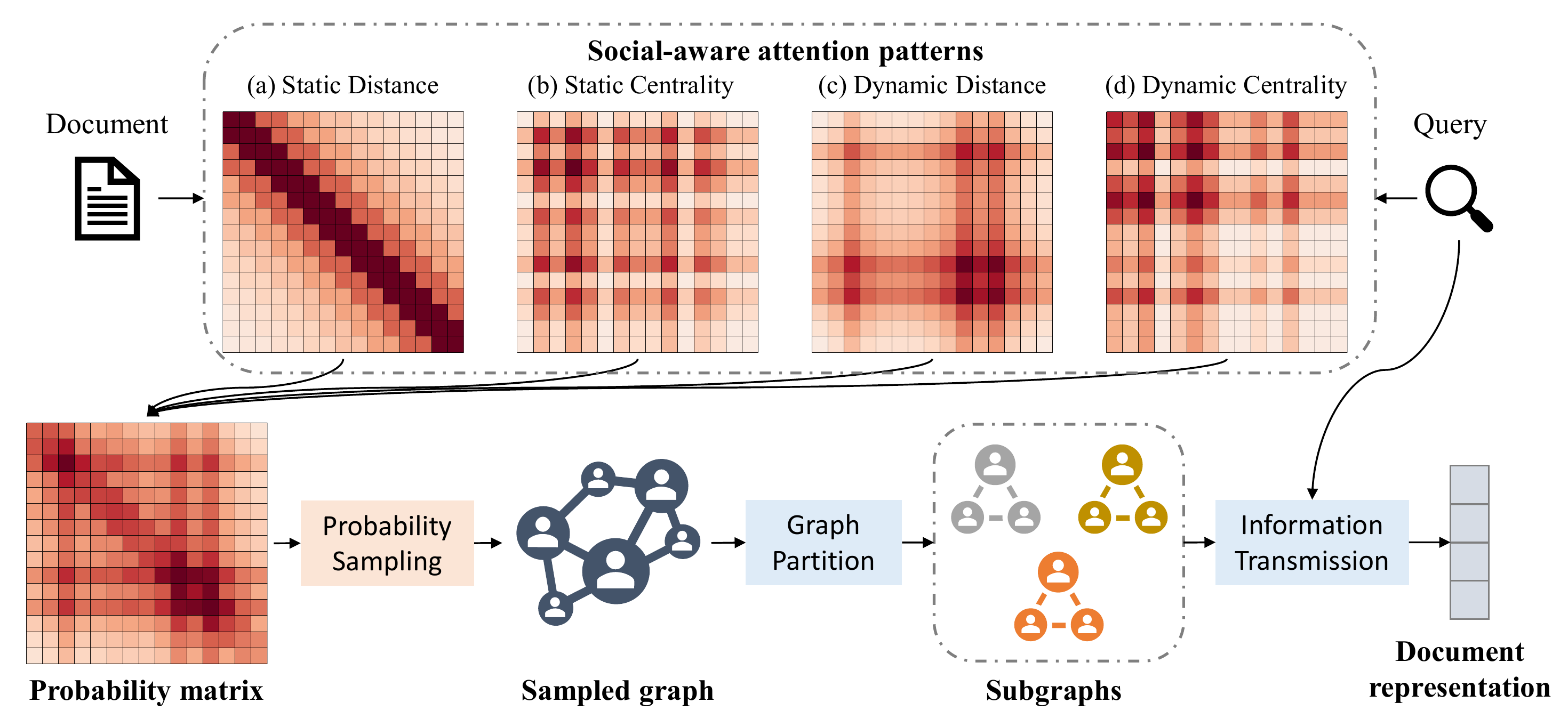}
	\caption{The overview of Socialformer. For a long document, we combine four social-aware attention patterns to sample a token-level graph. Darker color indicates higher probability. The graph partition module and the information transmission module are designed to facilitate calculation. Finally, the global document representation is obtained for ranking.}
	\label{fig:probability}
\end{figure*}
Document ranking has become an indispensable component in many search engines. Recently, BERT-based models are applied to encode documents with deeper text understanding. For longer document texts, previous studies devise various long-document transformers with sparse attention patterns to reduce the complexity. However, most of them only pay more attention to local connections of terms. Inspired by social networks, we argue that remote edges between terms are crucial for effective information transmission across the whole document. The overview of Socialformer is shown in Figure~\ref{fig:probability}. To build a graph with reasonable remote edges, we incorporate the characteristics of social networks considering the influence of word distance and word centrality. To facilitate calculations, we segment the whole graph into multiple subgraphs according to the characteristics of friend circles. Then, we design a two-stage information transmission method to simulate the information flow in social scenarios. In the remaining part of this section, we will introduce the details.

\subsection{Social Network based Graph Construction}
As we stated in Section~\ref{sec:intro}, the characteristics of social networks (\ie, randomness, distance-aware, and centrality) ensure that most pairs of nodes in the sparse graph can reach with a short path. Efficient information transmission and sparsity are in line with our needs of designing attention patterns. In this section, we will introduce how to combine social networks to construct a graph.

Inspired by the randomness of social networks, we abandon the traditional fixed attention patterns. Instead, we sample the edges according to the social-aware probability. This enables us to construct diverse social networks for documents. To calculate the probability, we take the word distance and word centrality into account. In addition to the static probability which is only related to the document, we also consider the dynamic probability in response to the specific query. In fact, facing different queries, the contribution of each word in the document should not be the same. As shown in Figure~\ref{fig:probability}, there are four social-aware attention patterns we designed to compute the probability matrix. They are:

\textbf{Static Distance.} In addition to local connections, \citet{watts1998collective} believed that remote edges are necessary for information transmission. They proposed Watts-Strogatz model to randomly sample remote edges, which is applied to long document modeling by BigBird \cite{bigbird}. However, \citet{kleinberg2000navigation} pointed out that the random strategy does not match the real social scenarios. They claimed the probability two people are connected usually follows the inverse square law with their distance. Inspired by this, we argue that this rule is also in line with the long document modeling. The further the distance between two words, the lower the probability that they have semantic dependence. Formally, given a document $d$ with length $l$, denoted as $d=\{t_1, \cdots, t_l\}$, The static distance based probability of establishing an edge between tokens $t_i$ and $t_j$ is:
\begin{equation}
\label{eq:dis}
    P_{\text{sd}}(i,j)=\frac{1}{(1+|i-j|/p)^2},
\end{equation}
where $p$ is a hyper-parameter to control the probability range and is set to 50 in experiments.

\textbf{Static Centrality.} In social networks, some celebrities usually have connections with more people and have greater influence. Similarly, each word in the document has a different contribution to expressing the semantics of the document. We attempt to extract the ``celebrities'' in the document and highlight their influence. We choose a common indicator, TF-IDF weights, to indicate the static centrality of each word, denoted as $\{w^{\text{sc}}_1, \cdots, w^{\text{sc}}_l\}$. The static centrality based probability $P_{\text{sc}}(i,j)$ is related to the weights of tokens $t_i$ and $t_j$. We have:
\begin{equation}
    P_{\text{sc}}(i,j)=f(w^{\text{sc}}_i \cdot w^{\text{sc}}_j),
\end{equation}
where the function $f(\cdot)$ is used to map the weight product to probability. It consists of a smoothing layer and a normalization layer:
\begin{equation}
\begin{split}
    s(i, j)&= \text{smooth}(w_i \cdot w_j), \\
    f(i, j)&=\frac{s(i, j)-\min s(i,j)}{\max s(i,j)-\min s(i,j)},
\end{split}
\end{equation}
where $\text{smooth}(\cdot)$ is the smoothing function, which is implemented by $\text{sqrt}(\cdot)$ in experiments. It can be replaced by more sophisticated methods in the future.

\textbf{Dynamic Distance.} Given the query $q$, we assume that query terms contained in the document are more critical for document modeling, and their surrounding words in the document are usually more informative for the query. Formally, we extract the words that exactly match the query in the document as a set, denoted as $\{t^q_1,t^q_2, \cdots, t^q_n\}$. The weight of $i$-th document word $w^{\text{dd}}_i$ is related to the distance to these head words. We have:
\begin{equation}
    w^{\text{dd}}_i=\frac{1}{n} \sum^n_{j=1} \frac{1}{1+|i-\text{pos}(t^q_j)|/p},
\end{equation}
where $\text{pos}(\cdot)$ is to compute the original position in the document, and $p$ is the same hyper-parameter as in Eq.~(\ref{eq:dis}). The computing of dynamic distance based probability is similar to above: 
\begin{equation}
    P_{\text{dd}}(i,j)=f(w^{\text{dd}}_i \cdot w^{\text{dd}}_j).
\end{equation}

\textbf{Dynamic Centrality.} Some infrequent words will play an important role in semantics when matching with the query. Concretely, BERT-based model has performed well on document ranking task, and attention weights at the special token '[CLS]' position can reflect the contribution of each word. However, due to length limitation, we cannot feed all the words of a long document into BERT. To handle this issue, we propose using simple model to select several relevant words and applying BERT model to compute accurate weights of them. At the first stage, cosine similarity is used to determine the relevance of each word in the document to the query. Then, we select top 512 relevant words and feed them into BERT model with following input format:
\begin{equation}
    \texttt{[CLS] query [SEP]} \texttt{ rel}_1 \texttt{ rel}_2 \cdots \texttt{ rel}_n \texttt{ [SEP]}.
\end{equation}
We replace the cosine similarity weights of relevant words with BERT weights. Similarly, the dynamic centrality based probability is related to the weight of each term:
\begin{equation}
    P_{\text{dc}}(i,j)=f(w^{\text{dc}}_i \cdot w^{\text{dc}}_j).
\end{equation}

Finally, based on these four strategies, we take the weighted average of the four probability matrices $P=\lambda_1 P_{\text{sd}}+\lambda_2 P_{\text{sc}}+\lambda_3 P_{\text{dd}}+\lambda_4 P_{\text{dc}}$ for sampling. To control the sparsity of generated graph, we set a hyper-parameter $\mu$ to scale probability following:
\begin{equation}
\label{eq:sparsity}
     \sum_{i,j}\frac{P_{ij}}{\mu} = l^2 * (1-sparsity),
\end{equation}
where $l$ is the max length of documents. The adjacency matrix $M$ of the graph is sampled on the scaled probability matrix $P$ by:
\begin{align*}
\begin{split}
M_{ij}= \left \{
\begin{array}{ll}
   1, & \text{if $\text{random}(0,1)<P_{ij}$;}\\
   0, & \text{otherwise,}
\end{array}
\right.
\end{split}
\end{align*}
where $\text{random(0,1)}$ means getting a random number from 0 to 1. However, due to the randomness of our sampling strategy, the edges of the sampled graph are unstructured. It is hard to compute the self-attention matrix like traditional attention patterns. To handle this issue, we attempt to divide the whole graph into multiple subgraphs while retaining as much information as possible.

\begin{figure}[!t]
	\centering
	\setlength{\abovecaptionskip}{0.1cm}
	\includegraphics[width=0.9\linewidth]{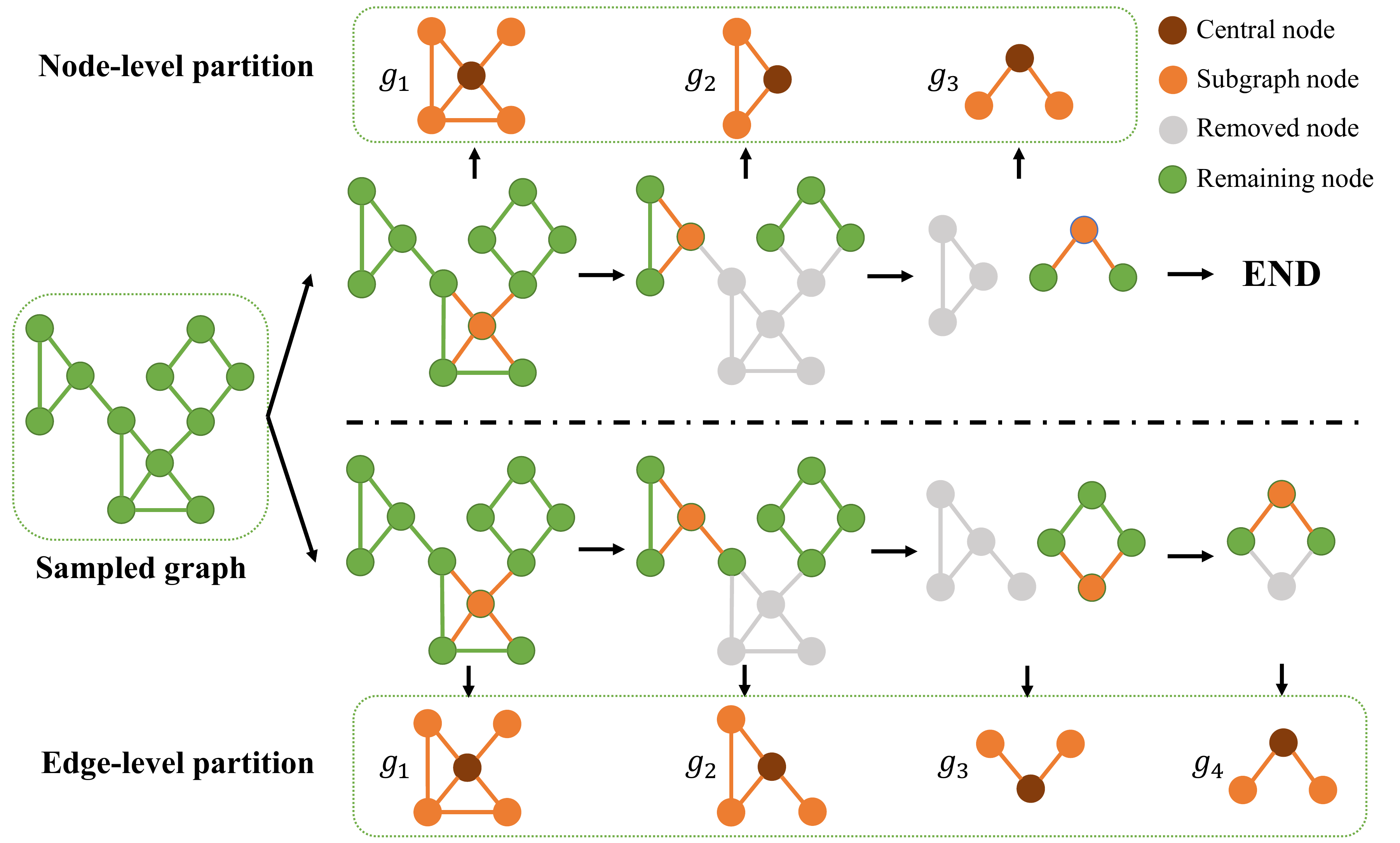}
	\caption{The overview of graph partition. Inspired by friend circles, node-level partition and edge-level partition are devised to segment the graph into multiple subgraphs.}
	\label{fig:partition}
\end{figure}
\subsection{Graph Partition}
The reason why the previous sparse attention patterns can reduce the complexity is that they can be easily split into multiple small self-attention blocks \cite{longformer, bigbird}. However, due to the randomness of our sampled graph, the edge distribution is unstructured. We propose to segment it into multiple subgraphs for calculation. We expect these subgraphs to retain as many nodes and edges as possible to minimize the loss of information. To determine the way of graph partition, we refer to another feature of social networks: the relationships between people are usually formed based on friend circles. In social scenarios, the friend circle is a common relationship structure, such as classmates and relatives. One of the characteristics of the friend circle is that there is often one person at the core who is responsible for connecting people in the entire circle~\cite{DBLP:conf/sigir/ZhouDWXW21}. This feature provides us with a way to extract friend circles.


\begin{figure*}[!t]
	\centering
	\vspace{-0.2cm}
	\setlength{\abovecaptionskip}{0.1cm}
   \setlength{\belowcaptionskip}{0.1cm}
	\includegraphics[width=0.8\linewidth]{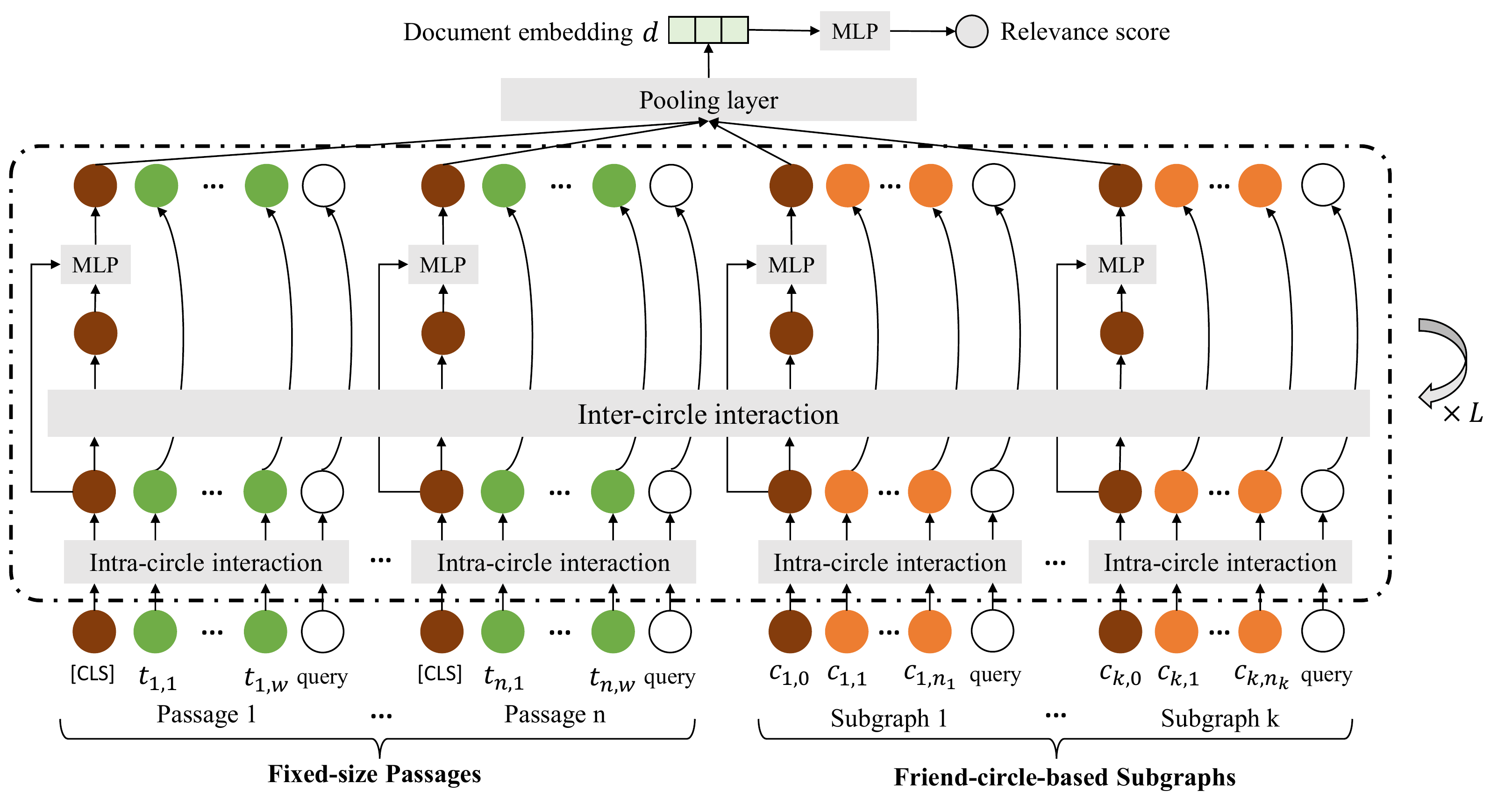}
	\caption{The architecture of information transmission model. Integrating fixed-size passages and social-aware subgraphs, intra-circle and inter-circle interactions are applied to enhance the information transmission. Finally, by aggregating the information of central nodes, we obtain the comprehensive document embedding to compute the relevance score.}
	\label{fig:model}
\end{figure*}


Specifically, we devise two partition strategies as shown in Figure~\ref{fig:partition}: node-level partition and edge-level partition. The former assumes that one node only appears in one subgraph, while the latter allows each node to belong to different subgraphs. In a limited number of subgraphs, node-level strategy can record more node information, while the edge-level strategy retains more edges.

Formally, given the whole graph $\mathcal{G}=\{\mathcal{N},\mathcal{E}\}$, where $\mathcal{N}$ is the set of nodes containing the document words and $\mathcal{E}$ represents connections between words, our goal is to find out top $k$ informational subgraphs. Specifically, we first select the node with the highest degree as the central node of the first subgraph. Then, the central node and its neighboring nodes form the first subgraph $g_1$. To ensure the distinction between different subgraphs, there are two strategies of partition. For node-level partition, we remove all nodes in the subgraph $g_1$ from $\mathcal{G}$ and repeat the above process to form other subgraphs. For edge-level partition, we only remove edges in $g_1$ from $\mathcal{G}$, and some nodes still have a chance to appear in other subgraphs. Finally, we obtain $k$ subgraphs $G=\{g_1,...g_k\}$, which will act on information transmission in the next section.

\subsection{Iterative Information Transmission}
In social networks, the connections within a friend circle are usually dense, which are called strong ties~\cite{henning1996strong, hu2019strong}. They contribute to person-level interactions and information transmitted through strong ties tends to be redundant. By contrast, weak ties \cite{wilson1998weak} have a greater impact on the group-level spread of information. To imitate this pattern in long documents, as shown in Figure~\ref{fig:model}, we devise a two-stage information transmission model over all subgraphs and passages, which consists of the intra-circle interaction and the inter-circle interaction. With $L$ iterative stacking blocks, the information between most pairs of nodes can be transmitted to learn a global document embedding. The details are introduced as follows.

\textbf{Intra-circle Interaction.}
People who belong to the same friend circle often have certain similarities. Triadic closure theory~\cite{easley2012networks} shows two people in the same circle have higher probability of becoming friends. Based on such observation, we use a fully-connected transformer layer to achieve information transmission in each subgraph. Formally, for the subgraph $g_i$, assuming it consists of a central node $c_{i,0}$ and $n_i$ neighboring nodes, \ie, $C_i=\{c_{i,0}, c_{i,1}, \cdots, c_{i,n_i}\}$, the intra-circle interaction with low-level transformer is defined as:
\begin{equation}
    \bC^{\text{low}}_i=\text{Trm}(\{c_{i,0}, c_{i,1}, \cdots c_{i,n_i}, \texttt{query}\}),
\end{equation}
where $\text{Trm}(\cdot)$ is the transformer encoder. The output of this layer is denoted as $\bC^{\text{low}}_i=\{\bc^{\text{low}}_{i,0}, \bc^{\text{low}}_{i,1}, \cdots, \bc^{\text{low}}_{i,n_i}\}$. In the remaining of this section, we use $\bc^{\text{low}}_i$ to denote the central node $\bc^{\text{low}}_{i,0}$, which represents this circle for inter-circle interaction.

\textbf{Inter-circle Interaction.}
Connections between different friend circles can help the information to be transmitted to further places. To promote the semantics of each word in the document to be transmitted to all positions, we design an inter-circle interaction layer over central nodes with a high-level transformer. To preserve the sentence structure information, we integrate fixed-size passages with subgraphs together for information transmission. Assuming that there are $m$ passages and $k$ subgraphs, we take the central nodes of each subgraph and passage (regarding ``[CLS]'' as the central node) as the input, \ie, $\bC^{\text{low}}=\{\bc^{\text{low}}_1, \cdots, \bc^{\text{low}}_{k+m}\}$. We have:
\begin{equation}
    \bC^{\text{high}}=\text{Trm}(\{\bc^{\text{low}}_1, \cdots, \bc^{\text{low}}_{k+m}\}).
\end{equation}
The output $\bC^{\text{high}}=\{\bc^{\text{high}}_1, \cdots, \bc^{\text{high}}_{k+m}\}$ considers the information transmission across all subgraphs and passages, and will take the global information to its neighboring nodes in the next iteration. 

\textbf{Iterative Stacking Blocks.}
In order to promote global information to be spread to every node, it is more reasonable to alternate the intra-circle and inter-circle interactions. The overall structure is composed of $L$ stacking blocks. Each block contains an intra-circle and an inter-circle interaction layer. The outputs of two transformer layers will be aggregated as the inputs to the next block:
\begin{align*}
\begin{split}
\{\bc_{i,j}\}_{L}= \left \{
\begin{array}{ll}
   [\{\bc^{\text{low}}_i, \bc^{\text{high}}_i\}_{L-1}]\cdot \bW^C, & \text{if $c_{i,j}$ is the central node;}\\
   \{\bc^{\text{low}}_{i,j}\}_{L-1}, & \text{otherwise,}
\end{array}
\right.
\end{split}
\end{align*}
where $\bW^C \in \mathbb{R}^{2E \times E}$ is the projection matrix. In the iteration of the $L$ layers, central nodes serve as the bridge for global information transmission. The whole process is highly consistent with the exchange of information in social networks. 

\textbf{Aggregation.}
After $L$ stacking blocks, we aggregate all passages and subgraphs to learn the document embedding with global information. Following prior works, such as PARADE~\cite{parade}, we aggregate the representations corresponding to central nodes by a pooling layer to get the document embedding $\bd$, defined as: 
\begin{equation}
    \bd=\text{Pooling}(\{\bc^{\text{high}}_1, \cdots, \bc^{\text{high}}_{k+m}\}_{L}),
\end{equation}
where $\text{Pooling}(\cdot)$ is the aggregation function, which can be implemented by Mean, Max, Attention, Transformer, etc. Since the document embedding has encoded the query information, we can directly compute the relevance score by feeding the document embedding into a linear layer:
\begin{equation}
    \text{score}(\bd)=\bv ^\mathrm{T} \bd,
\end{equation}
where $\bv \in \mathbb{R}^E$ is a linear function to project the document embedding into a scalar score.

\subsection{Training}
For each query $q$ and a group of documents $G_q$, we choose the listwise cross entropy as the loss function following~\cite{bertmaxp}:
\begin{equation}
    \mathcal{L}^q=-\text{log}\frac{\text{exp}(\text{score}(\bd^+))}{\sum_{\bd \in G_q} \text{exp}(\text{score}(\bd))}
\end{equation}
where $\bd^+$ is the document embedding of positive sample, and $\bd$ is the document representation for a document $\bd \in G_q$.

\section{Experimental Settings}\label{sec:exp}
\subsection{Datasets and Evaluation Metrics}
To prove the effectiveness of our proposed Socialformer, we conduct experiments on the 2019 TREC Deep Learning Track Document Collection~\cite{DBLP:journals/corr/abs-2003-07820}. This collection is a large-scale benchmark dataset for web document retrieval. It contains 3.2 million documents with a mean document length of 1,600 words. We conduct experiments on two representative query sets widely used in existing works.
\begin{itemize}[leftmargin=*]
\item \textbf{MS MARCO Document Ranking~(MS MARCO)}~\cite{msmarco_dataset}: It consists of 367 thousand training queries, and 5 thousand development queries for evaluation. The relevance is rated in 0/1.
\item \textbf{TREC 2019 Deep Learning Track~(TREC DL)}~\cite{trecdl_dataset}: It replaces the test queries in MS MARCO with a novel set of 43 queries. Although it is smaller than MS MARCO, it has more comprehensive notations with the relevance scored in 0/1/2/3.
\end{itemize}
We use the official metrics to evaluate the top-ranking results, such as MRR@100 and nDCG@10. Besides, we also report  MRR@10 and nDCG@100 for MS MARCO and TREC DL, respectively.

\subsection{Baselines}
We evaluate the performance of our approach by comparing it with three groups of methods for modeling long documents:

(1) \textit{Traditional IR Models}. \textbf{BM25}~\cite{robertson2009probabilistic} is a highly effective probabilistic retrieval model based on IDF-weighted counting. \textbf{QL}~\cite{DBLP:journals/sigir/ZhaiL17} is another famous model which measures the query likelihood of query with Dirichlet prior smoothing.

(2) \textit{Passage-based Models}.
These methods firstly split the long documents into multiple passages with the fixed-size window, then use the standard Transformer architecture to predict the relevance of each small passage. \textbf{BERT-FirstP}~\cite{bertmaxp} predicts the relevance of each passage with BERT model independently, and uses the score of the first passage to represent the relevance of the whole document. \textbf{BERT-MaxP}~\cite{bertmaxp} combines the independent score of each passage with a max-pooling layer to ensemble the global relevance information. \textbf{IDCM}~\cite{idcm} is an intra-document cascade ranking model with an efficient passage selection strategy. \textbf{PARADE}~\cite{parade} proposes strategies for aggregating representations of document’s passages into a global document embedding and computes the final score.

(3) \textit{Long-document Transformer Models}. These methods handle long document ranking by designing sparse attention patterns in Transformer. \textbf{Longformer}~\cite{longformer} combines a local windowed attention with a task motivated global attention. We experiment with its both variants, \ie, standard \textbf{Longformer} and $\bm{{\rm Longformer}_{\rm Global}}$ with global attention. \textbf{QDS-Transformer}~\cite{qdstransformer} designs IR-axiomatic structures in transformer self-attention. \textbf{BigBird}~\cite{bigbird} combines global attention, local attention and random attention together for building a universal framework of sequence encoders.

Our method, which is called \textbf{Socialformer}\footnote{The code is available on \url{https://github.com/smallporridge/Socialformer}}, combines the advantages of passage-based models and long-document transformer models. We use $\bm{{\rm Socialformer}_{\rm node}}$ and $\bm{{\rm Socialformer}_{\rm edge}}$ to represent the model with two different graph partition strategies.

\begin{table*}[t!]
    \centering
    \vspace{-0.1cm}
    \small
     \setlength{\abovecaptionskip}{0.1cm}
 \setlength{\belowcaptionskip}{0.1cm}
    \caption{Results of all models on two document ranking benchmarks. ``$\dagger$'' denotes the result is significantly better than other models from the same setting in t-test with $p \textless 0.05$ level. The best results are in \textbf{bold} and the second best results are \underline{underlined}.}
    \label{table_a}
    \begin{tabular}{clcccccc}
    \toprule
     \multirow{2}[2]{*}{Model Type} & \multirow{2}[2]{*}{Model Name} & \multirow{2}[2]{*}{\minitab[c]{Doc. \\ Length}}  & \multirow{2}[2]{*}{\minitab[c]{Window \\ Size}} &  \multicolumn{2}{c}{MS MARCO} & \multicolumn{2}{c}{TREC DL} \\
        \cmidrule(lr){5-6} \cmidrule(lr){7-8}
        & & & & MRR@100 & MRR@10 & nDCG@100 & nDCG@10\\
        \midrule
        \multirow{2}{*}{\minitab[c]{Traditional \\ IR Models}} &  BM25 & - & - & $0.2538$ & $0.2383$  & $0.4692$ & $0.5411$ \\
        &  QL  & - & - & $0.2457$ & $0.2295$  & $0.4644$ & $0.5370$ \\
        \midrule
        \multirow{5}{*}{\minitab[c]{Passage-based \\ Models}} &  BERT-FirstP & 512 & 512 & $0.4321$ & $0.4268$ & $0.4949$ & $0.6202$ \\
        &  BERT-MaxP & 512 & 128 & $0.4173$ & $0.4088$ & $0.4835$ & $0.6014$ \\
        &  BERT-MaxP & 2048 & 128 & $0.4326$ & $0.4272$ & $0.4952$ & $0.6215$ \\
        &  IDCM & 2048 & 128 & $0.4367$ & $0.4280$ & $0.4960$ & $0.6235$ \\
        &  PARADE & 2048 & 128 & $\underline{0.4386}$ & $\underline{0.4312}$ & $0.4975$ & $0.6280$ \\        
        \midrule
        \multirow{4}{*}{\minitab[c]{Long-Document \\ Transformer Models}} & Longformer & 2048 & 128 & $0.4263$ & $0.4192$ & $0.4942$ & $0.6208$ \\
        &  ${{\rm Longformer}_{\rm Global}}$ & 2048 & 128 & $0.4381$ & $0.4302$ & $0.4982$ & $0.6292$ \\
        &  QDS-Transformer & 2048 & 128 & $0.4379$ & $0.4300$ & $\underline{0.4988}$ & $0.6315$ \\
        &  BigBird & 2048 & 128 & $0.4385$ & $0.4311$ & $0.4985$ & $\underline{0.6318}$ \\
        \midrule
        \multirow{4}{*}{\minitab[c]{Our Models}} &  $\rm Socialformer_{\rm node}$ & 512 & 128 & $0.4290^\dagger$ & $0.4231^\dagger$ & $0.4902^\dagger$ & $0.6084^\dagger$ \\   
        & $\rm Socialformer_{\rm edge}$ & 512 & 128 & $0.4313^\dagger$ & $0.4258^\dagger$ & $0.4950^\dagger$ & $0.6212^\dagger$ \\ 
        & $\rm Socialformer_{\rm node}$ & 2048 & 128 & $0.4483^\dagger$ & $0.4402^\dagger$ & $0.5087^\dagger$ & $0.6534^\dagger$ \\ 
        & $\rm Socialformer_{\rm edge}$ & 2048 & 128 & $\bm{0.4490}^\dagger$ & $\bm{0.4411}^\dagger$ & $\bm{0.5119}^\dagger$ & $\bm{0.6615}^\dagger$ \\ 
    \bottomrule
    \end{tabular}
    \label{tab:overall}
\end{table*}

\subsection{Implementation Details}
We re-rank the documents from Top100 results retrieved by the advanced retrieval model ANCE~\cite{DBLP:journals/corr/abs-2007-00808}. During training, for each query, we choose positive samples and negative samples in a 1:7 ratio. Negative samples are randomly selected from the candidate documents. Considering the balance of time cost and effect, all models are trained for one epoch with a batch size of 8. We use AdamW~\cite{DBLP:conf/iclr/LoshchilovH19} to optimize the parameters with learning rate of 1e-5. For the model, we use the hyper-parameter $\mu$ to control the sparsity of the social graph at about 0.93 level by Eq. (\ref{eq:sparsity}). The document length and window size are set to 2048 and 128 for experiments, and larger window size does not bring more improvement~\cite{idcm, qdstransformer}. To control memory complexity, the max number of subgraphs $k$ is set to 16, which could retain critical nodes or edges for information transmission. We set the number of layers $L$ to 12 and intra-circle interaction layers are initialized by BERT-base model. Considering the time cost, the pooling layer is set to max pooling operation, which is also applied to the baseline model PARADE.

\section{Experimental Results}
\subsection{Overall Results}
Experimental results on the MS MARCO and the TREC-DL 2019 datasets are shown in Table~\ref{tab:overall}. Some observations are as follows.

(1) Among all models, our social-aware models outperform all baselines with the same settings in terms of all evaluation metrics. Compared with the best baseline models, our models have significant improvements in both datasets with paired t-test at $p \textless 0.05$ level. Concretely, for MS MARCO dataset, our best model $\rm Socialformer_{\rm edge}$ outperforms PARADE by over 2.37\% improvement on MRR@100, while the improvement over BigBird is 4.70\% on nDCG@10 for TREC DL dataset. These results indicate that introducing the characteristics of social networks into attention patterns can improve the ranking quality.

(2) Comparing different model types, we find that information transmission among passages is effective in learning global document representations. Specifically, PARADE, which aggregates the representations of all passages, outperforms score aggregation methods such as BERT-MaxP. This indicates that the aggregated document representation can alleviate the problem of lack of global information in document embeddings. Moreover, long-document transformer models devise different attention patterns to achieve information transmission between passages, which shows comparable performance. Our model Socialformer refers to the social networks when designing attention patterns, so as to achieve more effective information transmission within the document.

(3) Comparing different versions of Socialformer, it can be observed that longer document input (2048 vs. 512) brings an obvious improvement in results. This conclusion can also be drawn on BERT-MaxP. This reveals longer context contains more useful information to understand the semantics of documents. Moreover, the performances of $\rm Socialformer_{\rm node}$ and $\rm Socialformer_{\rm edge}$ are similar for 2048 document length, but when we limit the input length to 512, $\rm Socialformer_{\rm edge}$ demonstrates greater superiority. This indicates that when the number of nodes in the social graph is relatively small, edge-level partition retains much information.

In summary, the experimental results show that \textbf{introducing the characteristics of social networks into designing sparse attention patterns is conducive to refinement of document representations in long document modeling}.

\subsection{Effects of Social-aware Attention Patterns}
\begin{figure*}[!t]
\vspace{-0.2cm}
    \begin{subfigure}{.24\linewidth}
    \centering
    \includegraphics[width=\linewidth]{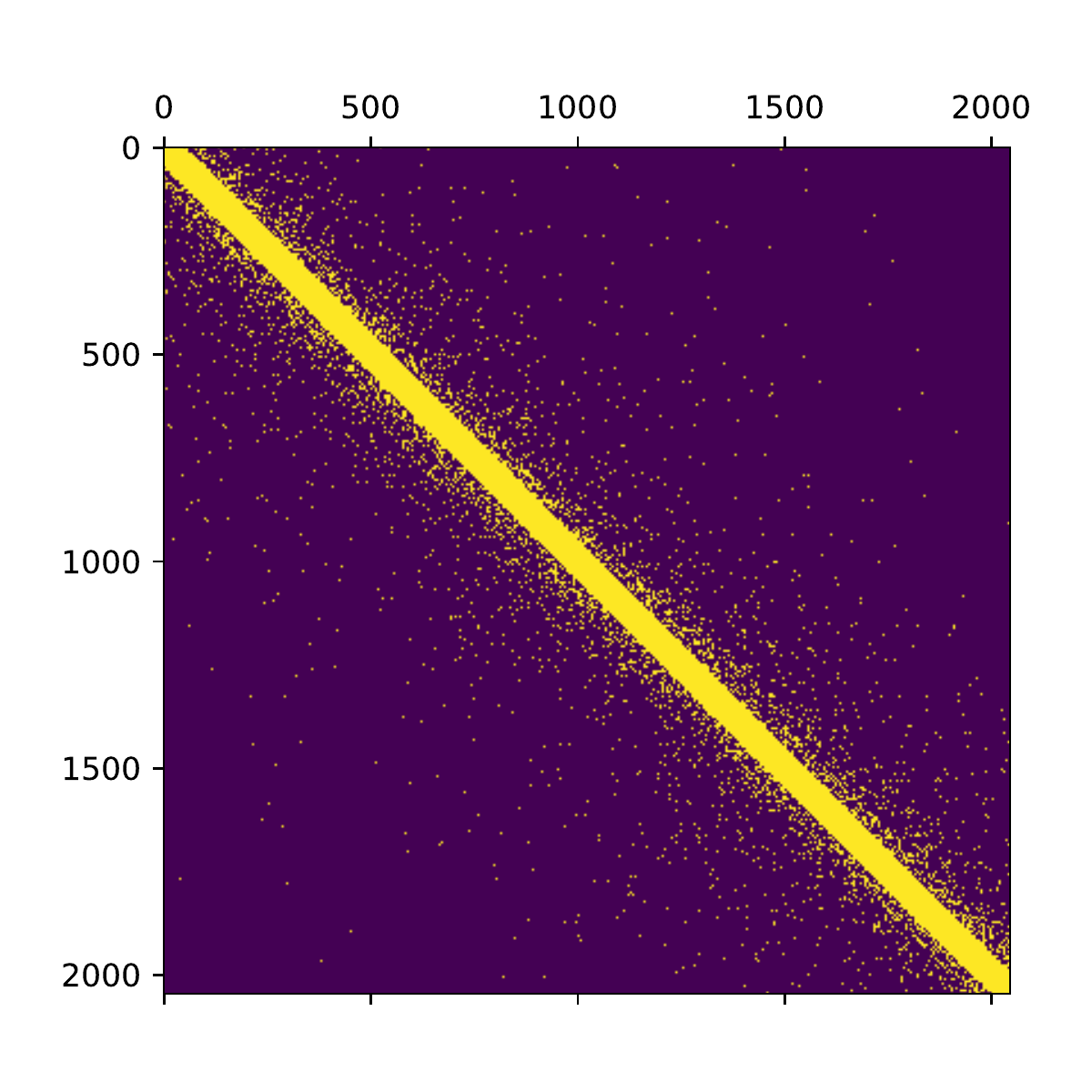}
    \caption{Static Distance}
    \end{subfigure}
    \begin{subfigure}{.24\linewidth}
    \centering
    \includegraphics[width=\linewidth]{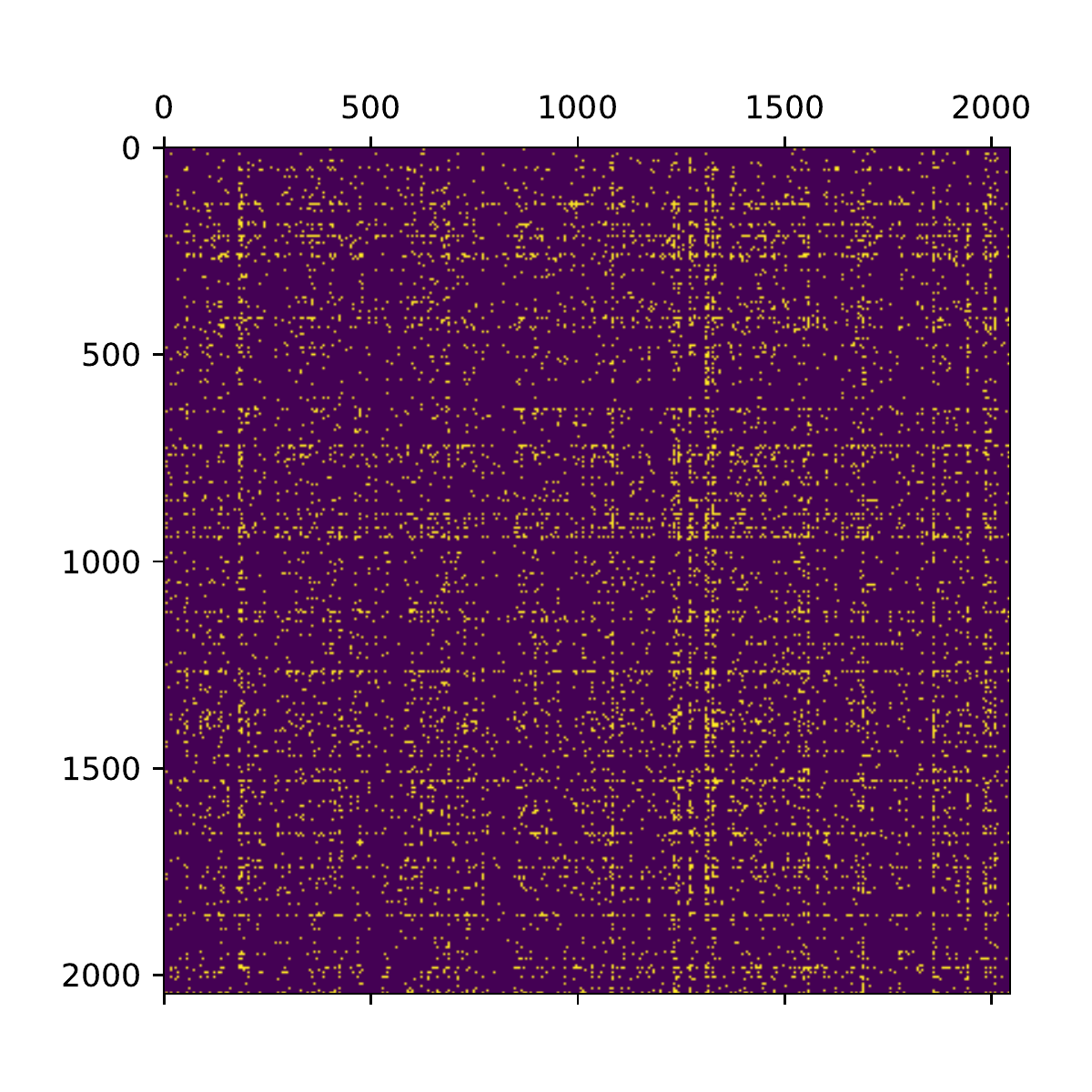}
    \caption{Static Centrality}
    \end{subfigure}
    \begin{subfigure}{.24\linewidth}
    \centering
    \includegraphics[width=\linewidth]{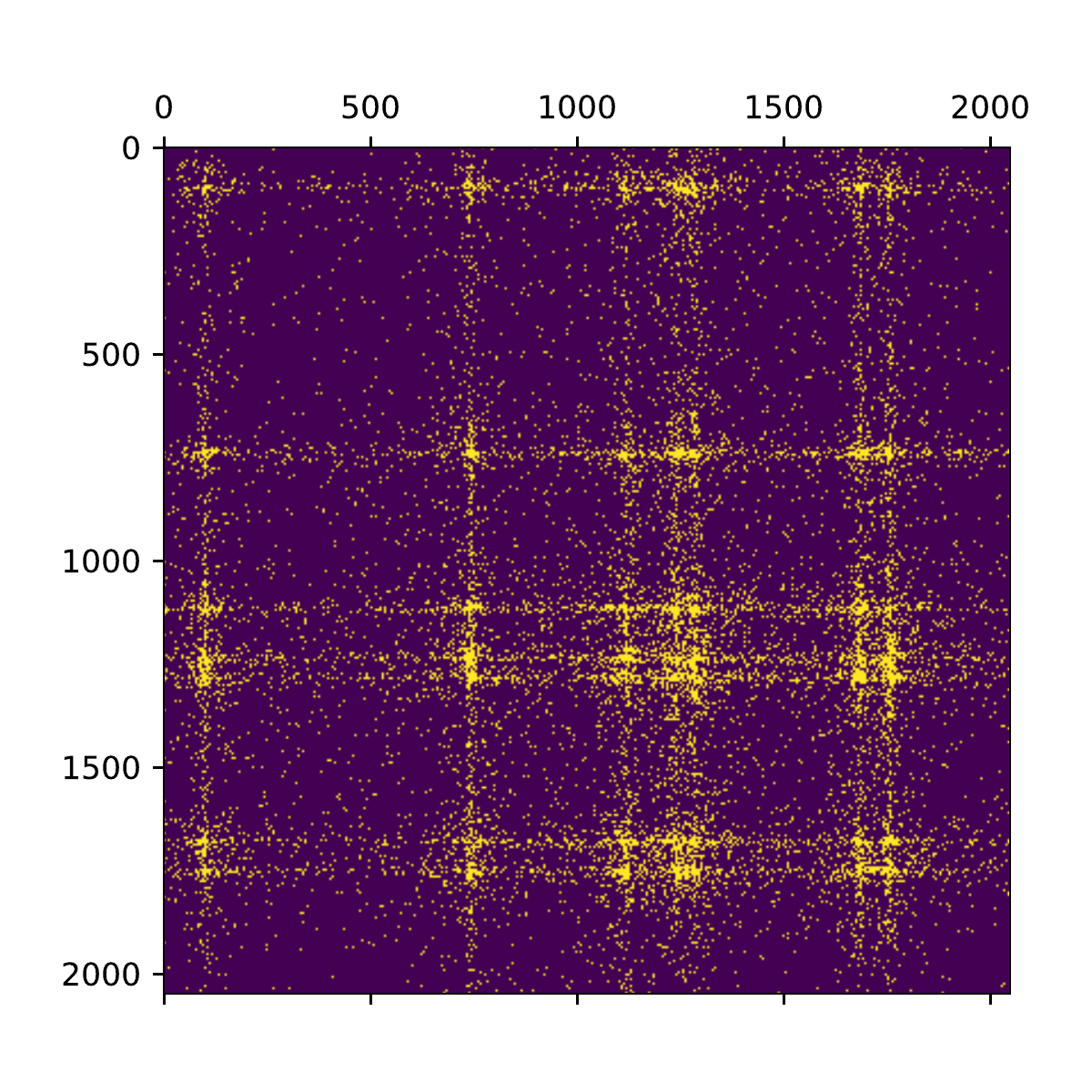}
    \caption{Dynamic Distance}
    \end{subfigure}
    \begin{subfigure}{.24\linewidth}
    \centering
    \includegraphics[width=\linewidth]{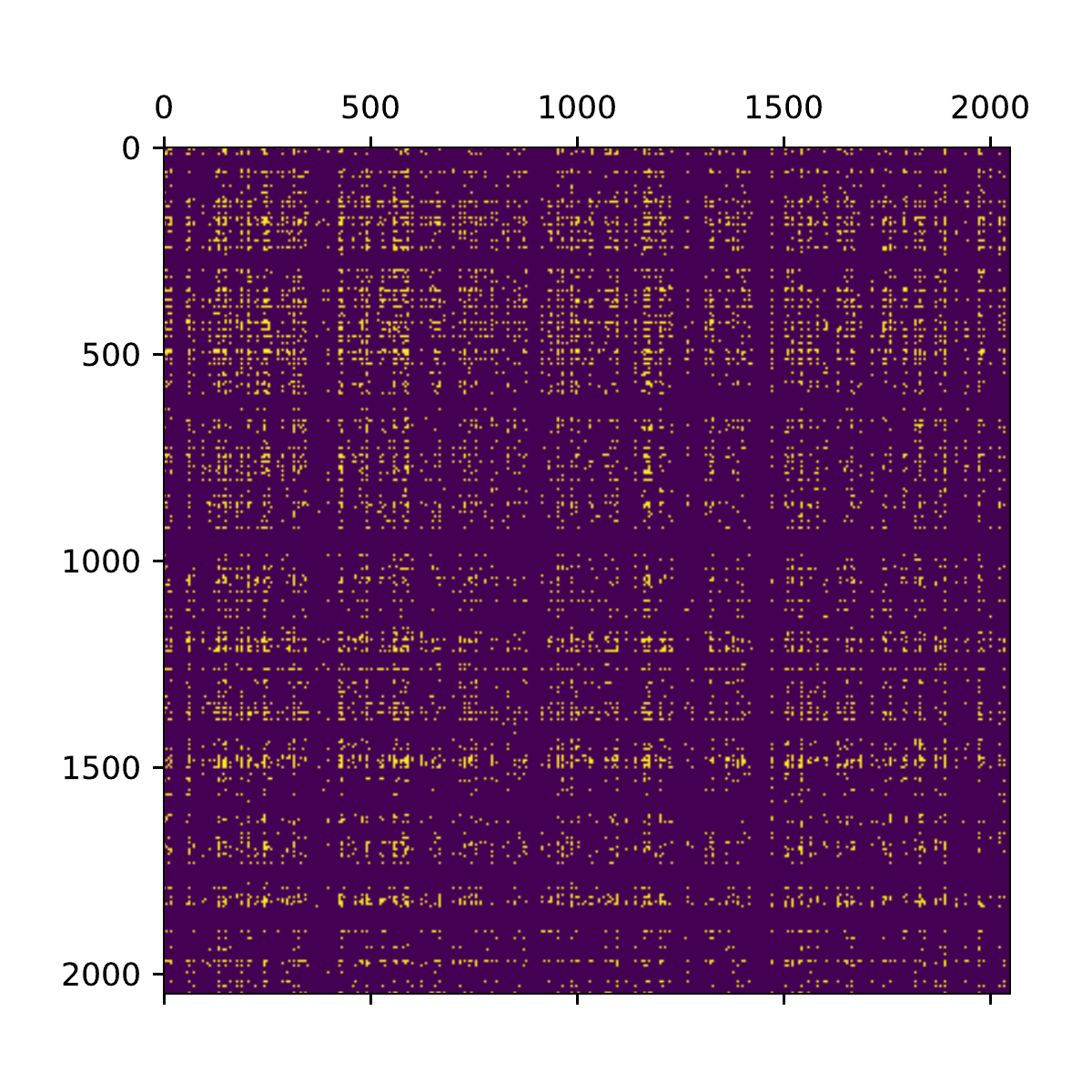}
    \caption{Dynamic Centrality}
    \end{subfigure}
    \caption{The adjacency matrix of using each attention pattern, and the yellow part means there is an edge.} 
    \label{fig:ajacency}
\end{figure*}
In the process of generating the graph, we calculate the probability matrix from the dynamic-static and distance-centrality dimensions respectively. To verify the necessity of each of our attention patterns, we explore the role of each strategy, including probability matrices of static distance, static centrality, dynamic distance, and dynamic centrality. In order to directly observe the effect of each attention pattern, we visualize the adjacency matrix generated by each strategy with 0.9 sparsity. As shown in Figure~\ref{fig:ajacency}, each strategy highlights different parts of the adjacency matrix to model relations between words. For further analysis, we remove one strategy at a time to observe the impact on the MS MARCO dataset. In addition, we use random sampling that the probability of establishing each edge is equal for comparison.

\begin{table}[!t]
 \center
 \small
 \setlength{\abovecaptionskip}{0.1cm}
 \setlength{\belowcaptionskip}{0.1cm}
 \caption{Performance of ablation studies of attention patterns on MS MARCO dataset.}
  \label{tab:ablation}
  \begin{tabular}{l|ll|ll}
  	\toprule
  	Model &\multicolumn{2}{c|}{MRR@100} & \multicolumn{2}{c}{MRR@10} \\ \hline
  	PARADE & 0.4382 & -2.41\% & 0.4302 & -2.47\% \\ \hline
  	$\rm Socialformer_{\rm edge}$ & 0.4490 & - & 0.4411 & - \\
  	\;\;w/o. static distance& 0.4469 & -0.47\% & 0.4380 & -0.70\% \\
  	\;\;w/o. static centrality & 0.4478 & -0.27\% & 0.4392 & -0.43\% \\
  	\;\;w/o. dynamic distance & 0.4447 & -0.96\% & 0.4359 & -1.18\% \\
  	\;\;w/o. dynamic centrality & 0.4450 & -0.89\% & 0.4364 & -1.06\% \\
	\;\;random edges & 0.4398 & -2.05\% & 0.4320 & -2.06\% \\
    \bottomrule
  \end{tabular}
\end{table}

The results are shown in Table~\ref{tab:ablation}. We find that the removal of each attention patterns will damage the results on all evaluation metrics. Concretely, deleting the dynamic patterns causes the most obvious impact on performance. This indicates that building the semantic dependencies of documents based on query is more helpful for learning global document representation. Meanwhile, the static patterns also make some contributions to the results. The four strategies work together to build a graph like social networks in the document. Additionally, using the strategy of randomly building edges instead of our attention patterns causes a severe drop on the results. This shows that using the characteristics of social networks can promote the transfer of information in documents. After removing the social features, the model mainly carries out two-stage information transmission through passages, which has similar performance to PARADE.

\subsection{The Effect of Sparsity on Graph Partition}
\begin{figure}[!t]
\setlength{\abovecaptionskip}{0.1cm}
\setlength{\belowcaptionskip}{0.1cm}
    \begin{subfigure}{.49\linewidth}
    \centering
    \includegraphics[width=\linewidth]{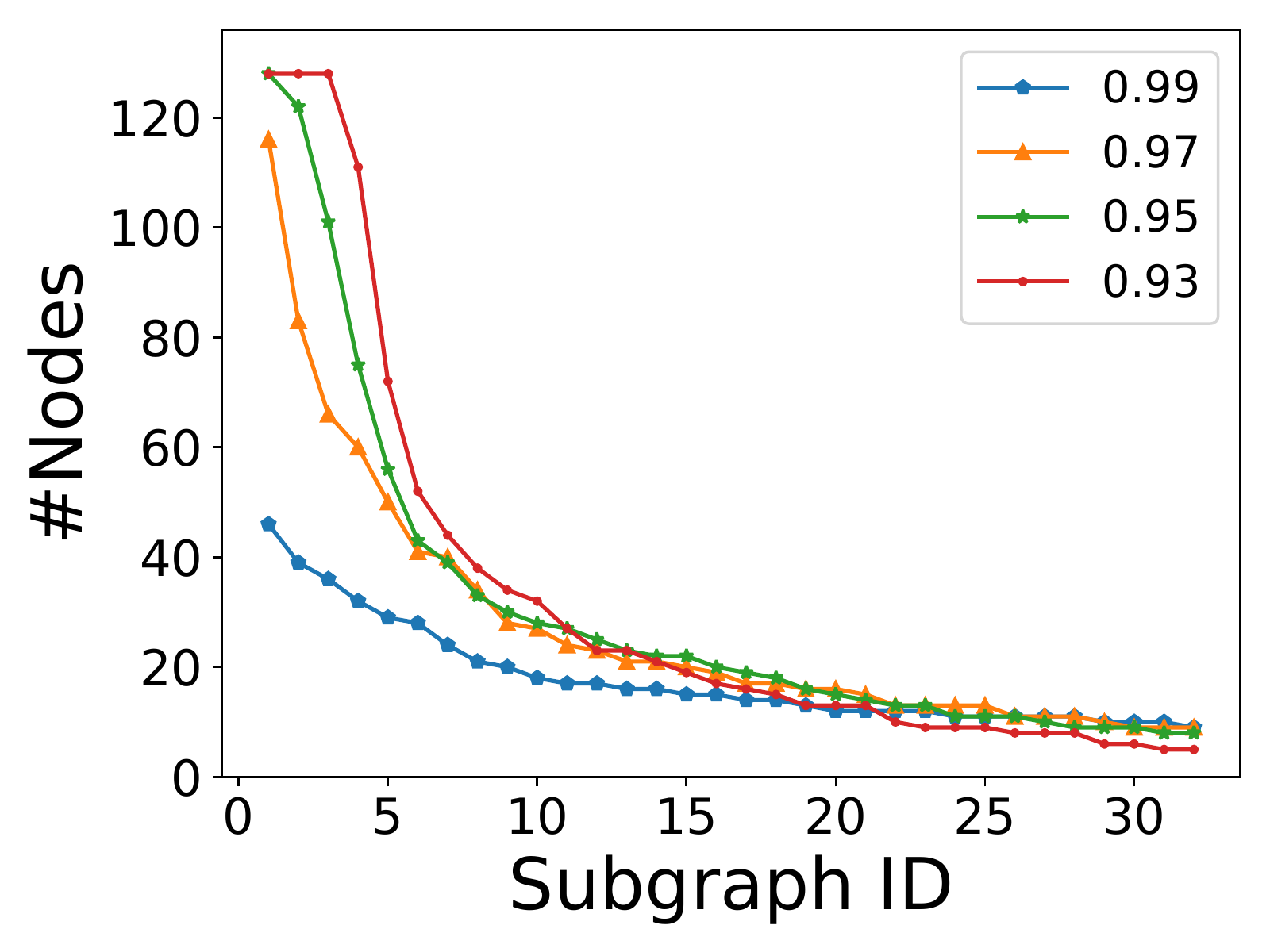}
    \caption{Node-level partition}
    \end{subfigure}
    \begin{subfigure}{.49\linewidth}
    \centering
    \includegraphics[width=\linewidth]{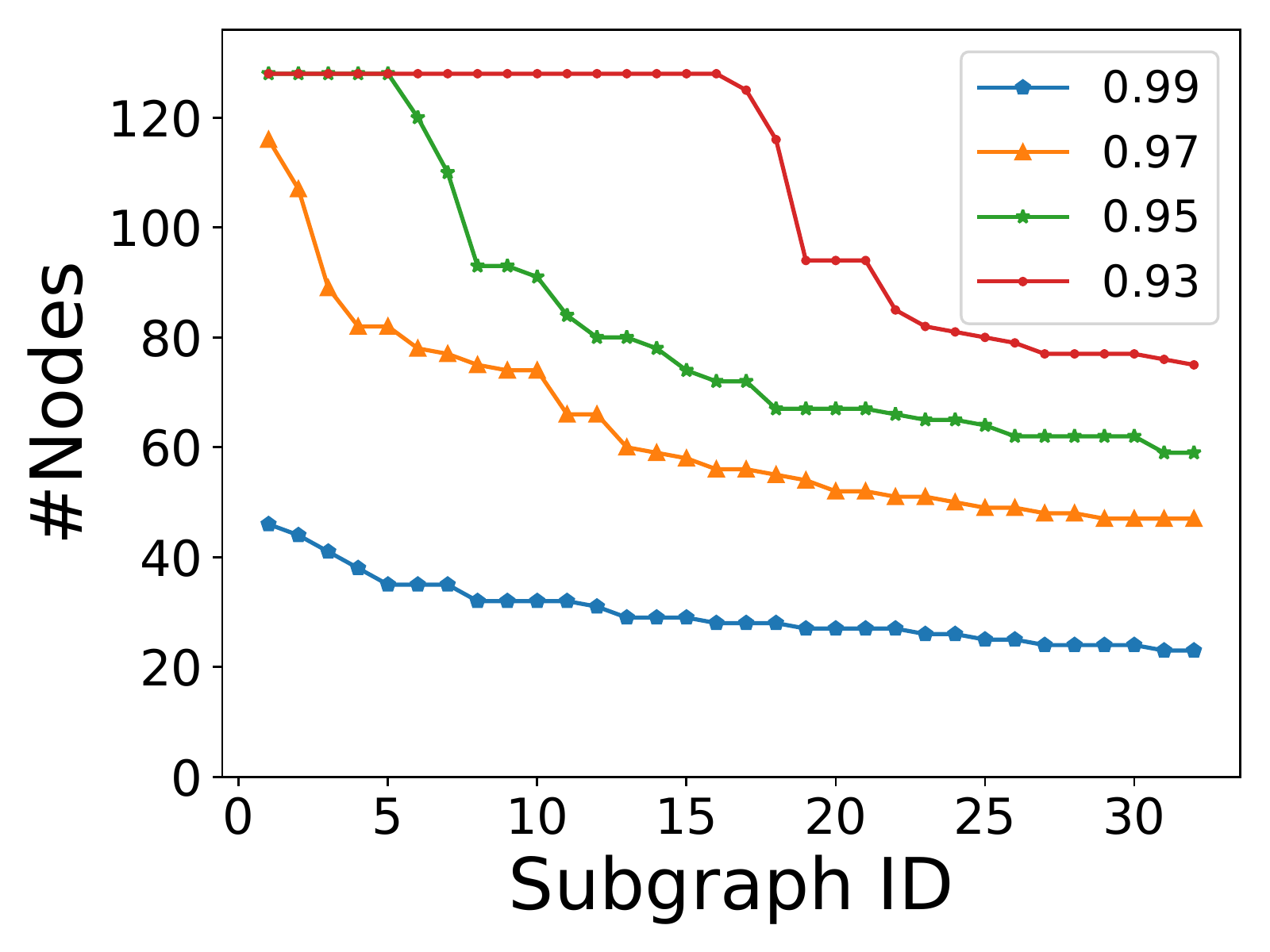}
    \caption{Edge-level partition}
    \end{subfigure}
    \caption{The relationship between the number of subgraph nodes and the sparsity of the graph.} 
    \label{fig:sparsity}
\end{figure}

Sparsity is an important hyper-parameter in the process of building the graph. Lower sparsity can enhance information transmission. However, it will also cause higher computational complexity, which leads to more subgraphs in graph partition. In order to compare the impact of different sparsity on graph partition, we select a document with 2,000 tokens, and set the sparsity of the graph at 0.99, 0.97, 0.95, 0.93 level respectively following Eq. (\ref{eq:sparsity}). We observe the relationship between the number of nodes (with maximum value of 128) of top 32 subgraphs and the sparsity. 

The results of two graph partition strategies are shown in Figure~\ref{fig:sparsity}. We observe that as the sparsity increases, the number of nodes in the top 32 subgraphs will also increase. When the sparsity reaches 0.93, the number of nodes in top 16 subgraphs of edge-level partition reaches the upper limit. Lower sparsity cannot bring more information if we set the max number of subgraphs to 16. This is why we choose 0.93 as the sparsity in experiments. Comparing the two strategies, we find that the number of nodes in node-level partition drops quickly. The reason is that there is no overlap between the nodes of each subgraph. In edge-level partition, more connections are retained, but there are many non-central nodes that cannot be included in top 32 subgraphs. In order to further explore the pros and cons of the two strategies, we explore the effect of different document lengths in the next section.

To observe what kind of query set the model is suitable for, we divide the whole query set on MS MARCO to four subsets based on the length $l$ of corresponding positive documents: (a) <512; (b) 512-1024; (c) 1024-2048; (d) >2048. We choose a baseline model PARADE and our two models for comparison.

\subsection{Experiment with Document Lengths}
\begin{figure}[!t]
	\centering
	\setlength{\abovecaptionskip}{0.1cm}
    \setlength{\belowcaptionskip}{0.1cm}
	\includegraphics[width=0.76\linewidth]{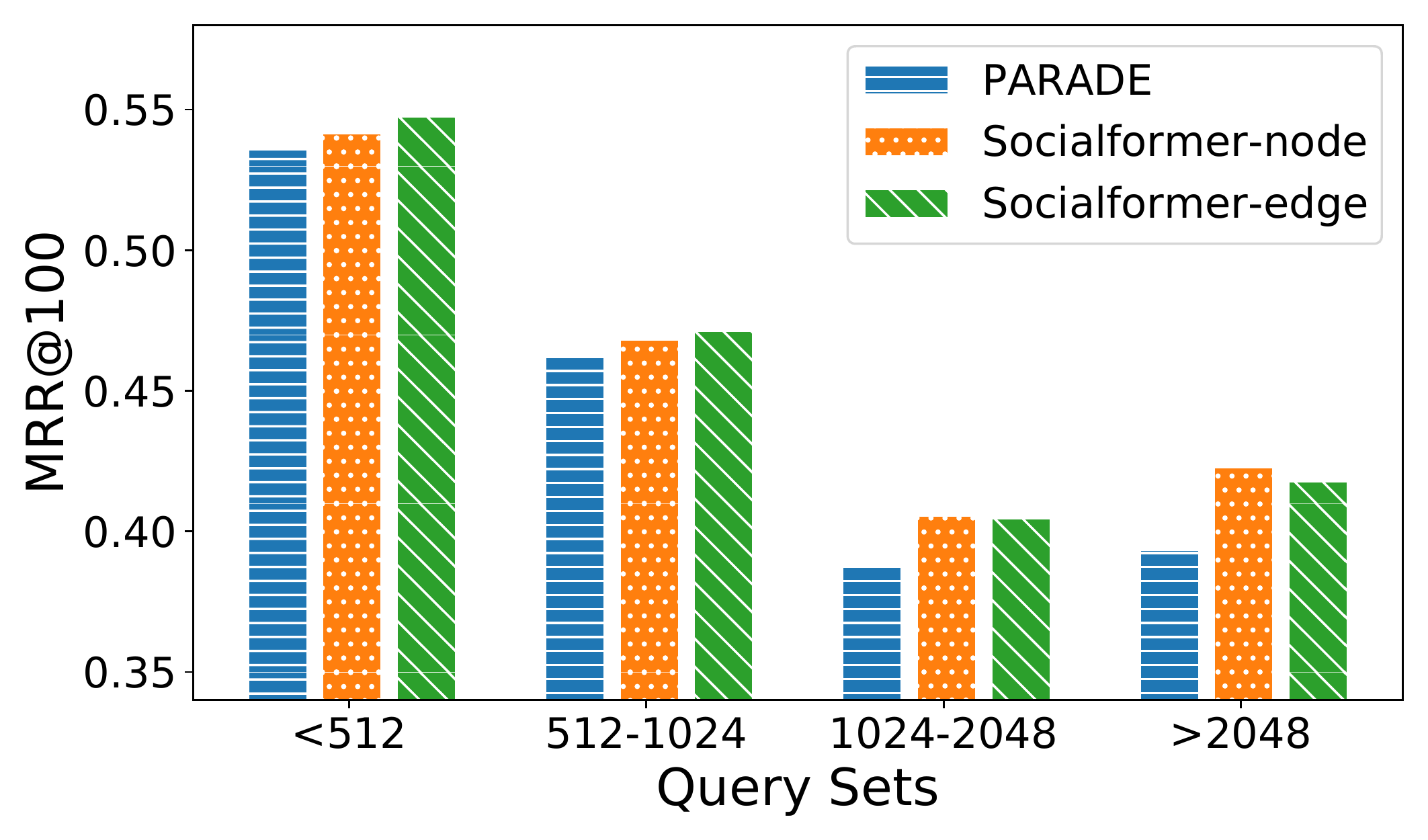}
	\caption{Performance with different query sets related to document length.}
	\label{fig:length}
\end{figure}
From Figure~\ref{fig:length}, we find that our social-aware models perform better than the baseline model on all query sets. Specifically, the gap between Socialformer and PARADE is widening as the document length grows. This indicates that building direct remote edges based on social networks enables the model to understand long documents better. Moreover, comparing two graph partition strategies, edge-level partition shows superiority when the document length is short, while the node-level partition performs better for longer texts. A possible reason is that top $k$ subgraphs of $\rm Socialformer_{\rm edge}$ can keep more edge information for short texts than $\rm Socialformer_{\rm node}$. When the document length grows, more node information is abandoned in top $k$ subgraphs. But for node-level partition, the majority of nodes can be retained regardless of the length of the document. 

\section{Conclusion}\label{sec:conclusion}
In this paper, we propose a social network inspired method for long document modeling. Concretely, we devise four attention patterns related to social networks and use probability sampling to construct a graph like social networks. To limit the computational complexity, the graph is divided into multiple subgraphs by two partition strategies. Then, to promote the full transmission of semantics in long documents, we present an iterative information transmission method which consists of inter-circle and intra-circle interactions. Finally, we can get a global document representation by an aggregation layer to re-rank the results. We conduct extensive experiments to verify the effectiveness of Socialformer. In the future, we will explore more sophisticated attention patterns and graph partition strategies according to features of webpage texts.

\begin{acks}
Thanks for reviewers' valuable comments. Zhicheng Dou is the corresponding author. This work was supported by National Natural Science Foundation of China No. 61872370, Beijing Outstanding Young Scientist Program NO. BJJWZYJH012019100020098, China Unicom Innovation Ecological Cooperation Plan, the Outstanding Innovative Talents Cultivation Funded Programs 2020 of Renmin University of China, and Intelligent Social Governance Platform, Major Innovation \& Planning Interdisciplinary Platform for the ``Double-First Class'' Initiative, Renmin University of China. We also acknowledge the support provided and contribution made by Public Policy and Decision-making Research Lab of RUC.
\end{acks}

\newpage
\balance
\bibliographystyle{ACM-Reference-Format}
\bibliography{sample-base}

\end{document}